# Wave vector star channel and star channel group in the reciprocal lattice space of crystal

Il Hwan Kim[a], Jong Ok Pak[b], Il Hun Kim[a, c], Song Won Kim[a], Lin Li[c]

[a] Department of physics, Kim Hyong Jik Normal University, Pyongyang, Democratic People's Republic of Korea

[b] Department of physics, Pyongyang University of Mechanical engineering, Pyongyang, Democratic People's Republic of Korea

[c] College of Science, Northeastern University, Shenyang 110819, People's Republic of China

Corresponding author: ririn@sohu.com (L. Li)

**Abstract** In the paper, a new method determining the wave vector star channel in the reciprocal lattice space of crystal in the light of the translational symmetry breaking is proposed, and, in order to consider the phase transitions according to the wave vector star channel, the conception of wave vector star channel group is adopted. By this method, it is revealed that the phase transitions in crystal are induced not by any arbitrary combinations of arms of the given star, but by the selected combinations of arms which are satisfied by symmetry of the parent phase. The wave vector star channel group is defined as the set of elements of space group leaving the wave vector star channel invariant. We show that the conception of wave vector star channel group can be efficiently used in studying the translational symmetry breaking related to all the Lifshitz wave vector stars of 230 space groups.

1. Introduction

For developing new materials, e. g., high temperature superconductor, ferroics such as ferroelectrics, ferromagnetics and ferroelastics, multi-ferroics such as ferroelectric-ferromagnetics and so on, and for studying those features, the study of symmetry is especially important, because the interesting physical properties are related with the breaking of rotational and translational symmetry of crystals and are described by the full irreducible representations of 230 space groups.

The translational symmetry breaking phenomena in crystals were first studied by Lifshitz in 1941.[1] Lifshitz studied systematically all the possible Bravais lattice types induced by the irreducible representation of space group, when the second-order phase transition occurs in crystal. Thereafter, in 1976, in order to explain not only the second-order phase transition but also "the first-order phase transition near the second-order", Naish[2, 3] first defined the wave vector star channel, being also called "transition channel" or "phase transition channel", and, using the conception, concretely classified Bravais lattices[1] suggested by Lifshitz.

The conception of the star channel has been used widely for studying phase transition in crystal. But in Ref. [3], it only indicated some of lattice types, with relatively high symmetry, which are obtained through phase transition. And so, in 1983, Jaric[11] defined the wave vector substar as all possible combinations of arms of wave vector star, and proposed the method finding out the space group of lower symmetry phase by using the conception.

But, applying the conception of the star channel to study phase transition of crystal, we can see that the previous results[1-3, 11] related with translation symmetry breaking are insufficient. In Ref. [3], the star channel was defined as a set of arms of star taking part in phase transition or as Bravais lattice types determined by them. In Refs. [2, 3, 11] it regarded the star channel as all the possible combinations of arms of wave vector star satisfying the condition of translational symmetry, $e^{-i\mathbf{k}(\alpha)t_D} = 1$, i.e., star channel and substar being essentially equal. But, among the wave vector star channels shown in Refs. [2, 3], there are some "unnecessary" channels not to be permitted by the translational symmetry. Existence of unnecessary channel makes a scientific mistake in studying phase transition.[12, 13]

In Refs. [12, 13], the cubic $O_h^1(Pm\bar{3}m)$ to tetragonal $D_{4h}^{17}(I4/mmm)$ improper ferroelastic first-order phase transition in $RAg_{1-x}In_x$ (where R=La, Ce, Pr) was discussed, and as a result, it indicated that the phase transition is induced by the $M_5$-brillouin zone boundary phonon in the CsCl structure, resulting in size change of 4 times in unit cell, and two arms of the wave vector star at the M point take part in the phase transition. But, in Ref. [14], it showed that the star channels related to two arms of the star are not permitted by symmetry, and therefore, in Refs. [12, 13] it was based on the incorrect model in which symmetry was not considered correctly. And it also showed that this is directly associated with limitation of method determining star channels.

In fact, if the wave vector star channel is correctly determined, we can know the particular arms of star taking part in phase transition and therefore, the wave vector star channel is very available in studying symmetry of lower symmetry phase. But, until now, in previous researches no attention have been paid to study on the basis of the particular arms of wave vector star taking part in phase transition.

In Refs. [15-17], the structural phase transition in $CsPbCl_3$ crystal, $O_h^1 \xrightarrow{319K} D_{4h}^5 \xrightarrow{317K} D_{2h}^{17} \xrightarrow{310K} D_{2h}^{16}(C_{2h}^2)$, was discussed. $CsPbCl_3$ crystal has $O_h^1$ of simple cubic lattice as the space group of high symmetry phase, and the phase transition is described by six-componential order parameter transformed according to the reducible representation of Lifshitz stars $k_{11}$ and $k_{13}$. Thus, the corresponding thermodynamic potential consists of the complicated basis with many invariants. So, in Ref. [17], the efficient potential model consisting of three-componential reducible order parameter was proposed on the basis of the fact that the number of non-zero components among six components of order parameter is three. In fact, in Ref. [17] the efficient potential characterizing the actual symmetry of lower symmetry phases was used, while the previous researchers have still used the six-dimensional reducible representation with higher symmetry than actual one for describing the phase transition.[41] After all, it results in a contradiction between group representation theory and phenomenological theory. [18, 19]

In Refs. [20, 21], in order to explain physical properties of $Pb_{1-x}Ca_xTiO_3$(PCT) crystal associated with complicated phase transitions according to x in the range 0<x<0.62, nine-componential order parameter model was suggested, which is transformed according to the reducible representation related to the wave vectors of R, M, Γ points in Brillouin Zone of space group $O_h^1$. In fact, due to high nonlinearity, it is very difficult to analyze the given experimental data such as phase diagram on the basis of nine-componential order parameter model. And so, in Refs. [20, 21], the fourth order thermodynamic potential model was used, and as a result, space groups of some phases were not determined.

The symmetry study[1~10, 24~32] of the structural phase transition based on the theory[23] of the second-order phase transition which was already suggested by Landau in 1930s, has been developing in the view not only of fundamentals but also of completing mathematical tools through old historical process. But, all previous methods are based on the full irreducible representation and the reducible representation of space group considering all arms of wave vector star.

Some researchers have attempted to consider directly the actual symmetry breaking by using the wave vector group and its representation related to one arm of the reciprocal lattice space in solving a series of problems.[33-36] Recently, in the field of condensed matter physics, the research for properties of the lower–dimensional systems with the broken symmetry has been attracting public attention, and the subperiodic groups have been used.[37-39] For example, there are 75 rod-groups and 80 layer groups, 17 plane-groups, 7 frieze–groups, line group and so on. These researches showed that the phase transition can be considered by the star channel.

The outline of this paper is as follows.

In Sec. 2, we are going to study the wave vector star channel characterizing the translational symmetry breaking of crystal in close relation with order parameter, unlike the preceding studies, and propose a kind of the new method for obtaining it. In Sec. 3, we are going to establish the theory of wave vector star channel group and show the applied examples of it.

**2. Method finding out the wave vector star channel**

By the Landau theory[23], when phase transition is induced by the certain irreducible representation of crystal symmetry group $G_0$, the change $\delta\rho$ in the distribution density function of lower symmetry phase is expanded in basis functions of the irreducible representation, $\varphi_l^{(m)}(\boldsymbol{r})$, as follows:

$$\delta\rho(\boldsymbol{r}) = \sum_{m,l} \eta_l^{(m)} \varphi_l^{(m)}(\boldsymbol{r}), \tag{1}$$

where $m$ is the number of arms of the wave vector star, $l$ is the number of basis functions of the full irreducible representation.

Because phase transition is related to each arm of wave vector star, we can write Eq. (1) simply as follows.

$$\delta\rho(\boldsymbol{r}) = \sum_{\alpha=1}^{m} \eta^{(\alpha)} \varphi^{(\alpha)}(\boldsymbol{r}) = \sum_{\alpha=1}^{m} \delta\rho_\alpha(\boldsymbol{r}) \tag{2}$$

From the condition of translational symmetry, $\hat{t}_D \delta\rho(\boldsymbol{r}) = \delta\rho(\boldsymbol{r})$, the following equation is obtained.

$$e^{-ik(\alpha)t_D} = 1, \quad (\alpha = 1 \sim m) \tag{3}$$

After all, the crystal lattice of lower symmetry phase is characterized by translation vector $t_D$ satisfying Eq. (3) with respect to the given wave vector star.

In this paper, we accepted the new space consisted by arms of star, by letting the arms of wave vector star and the order parameter one-to-one correspondence. In Eq. (2), $\varphi^{(\alpha)}(\boldsymbol{r})$ is the basis function for the representation of translation group, $T$, and therefore, $\eta^{(\alpha)}$ are regarded as components of order parameter characterizing the translational symmetry breaking. Because several arms of star can take part together in structural phase transition, $\{\eta^{(\alpha)}\}$ becomes multi-componential order parameter usually and is changed according to the representation $\Gamma(t)$ of translation group $T$.

Now introduce the representation space $\varepsilon_m$ spanned by $\Gamma(t)$ and establish the orthogonal coordinate system with components $\eta^{(\alpha)}$ in this space. We call this m-dimensional Euclidean space $\varepsilon_m$ the space of order parameter related with arms of wave vector star taking part in the translational symmetry breaking. Then, in this space, components $\eta^{(\alpha)}$ are the projections into coordinate axes $e_1, e_2, \cdots, e_m$ of the multidimensional vector $\vec{\eta}$.

Meanwhile, infinite group $T$ and it's representation spanned by $\varepsilon_m$-space could be corresponded to finite group $T_f = T/\ker\Gamma$ and faithful representation $\Gamma_f$. Because $\Gamma_f$ is an Unitary representation, the set of the matrices of $\Gamma_f$ in $\varepsilon_m$-space forms a certain multidimensional abstract point group, and so, we call this group the translation $L$-group. The translation $L$-group is isomorphic to image $\mathrm{Im}\Gamma(t)$ of the representation $\Gamma(t)$ of the translation group. The order of translation $L$-group is as follows.

$$[L_T] = [\Gamma_f] = \left[ T \Big/ \ker\Gamma(t) \right] \tag{4}$$

Then the wave vector star channel can be found out by a method finding out the non-zero invariant subspace of order parameter space which is invariant to all subgroups of the translation $L$-group.

$$H\vec{\eta} = \vec{\eta} \tag{5}$$

That is, by the vector $\vec{\eta}$ of Eq. (5), the wave vector star channels are determined.

The dimension of invariant subspace can be found out as follows:

$$r = \frac{1}{[H]} \sum_i \chi(h_i), \tag{6}$$

where $[H]$ is the order of subgroup of the translation group, $\chi(h_i)$ is the sum of characters of the representations with respect to all the elements of subgroup and $r$ is the number of arms of the wave vector star channel. This dimension just corresponds to the number of arms of wave vector star taking part in phase transition.

The vectors $\vec{\eta}$ are elements of different orbits decomposed by the translation $L$-group in the order parameter space. All the representatives for the set of orbits with the same structure consist of the set of invariant vectors of the translation $L$-group.

Thus, we can find out the star channels corresponding to invariant vectors of the translation $L$-group. The arms contained in the channel are just the ones of star taking part in phase transition.

We are going to denote the wave vector star channel $[k] = [k_i^{(\alpha)}]$ ($\alpha = 1 \sim m$), where $i$ is a star number according to the paper [40] and $\alpha$ is an arm number in the star channel. When some arms of the given star take part in phase transition, we are going to denote the wave vector star channel $[k] = [k_i^{(\cdots ijk\cdots)}]$.

We all found out 179 wave vector star channels associated with 80 Lifshitz stars of 230 space groups. We showed the result in Table 1, in contrast to Ref. [3], where symbol (*) denotes the star channel not to be permitted during the spontaneous symmetry breaking and $n$ means the volume change of unite cell. The expression of symbols follows to Ref. [40].

As shown, some star channels permitted in previous papers are not to be permitted according to our result. This means that not any arbitrary combinations of arms of the given star, but only the selected combinations of arms which is satisfied by symmetry of the parent phase relate with the translational symmetry breaking in crystal.

## 3. Wave vector star channel group

We are going to accept the conception of new group to consider the phase transition with the wave vector star channel.

If any symmetry breaking is related to the wave vector star channel $[k]$, we can write the change $\delta\rho$ in the distribution density function like Eq. (1).

$$\delta\rho_{[k]} = \sum_{k,l} \eta_l^k \varphi_l^k \qquad (7)$$

Where $\varphi_l^k$ are the basis functions related to some arms of star taking part in phase transition among basis functions of the irreducible representation of the space group shown in Eq. (1). And so, $\varphi_l^k$ are not related to the full irreducible representation. In fact, the symmetry reflected in Eq. (7) is essentially different from one of Eq. (1). As a result, the number of components of the order parameter $\eta_\beta = (\eta_1, \eta_2, \ldots, \eta_l)\,(m \geq l)$ found out by Eq. (7) becomes much less than one of the order parameter $\eta_\alpha = (\eta_1, \eta_2, \ldots, \eta_m)$ found out by Eq. (1)

If elements $\{g\}$ of the space group act on $\delta\rho_{[k]}$ of Eq. (7), Bloch functions $\varphi^k(r)$ are transformed one another between those corresponding to arms of the star channel, and then, arms of the star channel are transformed one another by not all the elements of the space group, but only some among those. We can newly call the set of elements of space group $G_0$ leaving the wave vector star channel invariant. We will call it the wave vector star channel group(or simply call it channel group) and expressed to $G_{[k]}$.

$$G_{[k]} = \{g_n \mid g_n[k] = [k] + b\} \qquad (8)$$

or

$$G_{[k_i^{(\cdots ijk\cdots)}]} = \{g_n \mid g_n[k_i^{(\cdots ijk\cdots)}] = [\cdots, k_i + b, k_j + b, k_k + b, \cdots]\} \qquad (9)$$

When we look at the point of view of the channel group $G_{[k]}$, the basis functions $\{\varphi_l^k\}$ in Eq. (7) form the basis of the full irreducible representation of $G_{[k]}$.

We can see that the star channel group is the group which the wave vector group is generalized about the star channel. And so, between the space group related to all arms of star, the star channel group and the little group(the wave vector group), there exists a group-subgroup relationship.

When we know all little groups $G_{k_i}$ related to each arm of the star, the star channel group can be written as follows:

$$G_{[k]} = \{\cdots \cap G_{k_i} \cap G_{k_j} \cap G_{k_k} \cap \cdots\} \cup \{g_l\}, \qquad (10)$$

Where the first term $\{\cdots \cap G_{k_i} \cap G_{k_j} \cap G_{k_k} \cap \cdots\}$ of Eq. (10) is corresponded to a unit element of the permutation group and the second term $\{g_l\}$ is corresponded to $n!-1$ elements of the permutation group except a unit element. That is,

$$\begin{pmatrix} k_1, & k_2, & \cdots, & k_n \\ k_1, & k_2, & \cdots, & k_n \end{pmatrix} = (e) = \{\cdots \cap G_{k_i} \cap G_{k_j} \cap G_{k_k} \cap \cdots\} \qquad (11)$$

$$\begin{pmatrix} k_1, & k_2, & \cdots, & k_n \\ k_i, & k_j, & \cdots, & k_l \end{pmatrix} = (i, j, \cdots l) = \begin{cases} g_l k_1 = k_i \\ g_l k_2 = k_j \\ \vdots \\ g_l k_n = k_l \end{cases} = g_l \qquad (12)$$

$$(i, j, \cdots l, = \overline{n})$$

In the view of the mathematics, $n!$ permutations consist of the permutation group $S_n(k_1, k_2, \cdots, k_n) \equiv S_{k_n}$.

$$\begin{pmatrix} k_1, & k_2, & \cdots, & k_n \\ P_{k_1}, & P_{k_2}, & \cdots, & P_{k_n} \end{pmatrix} = \begin{pmatrix} k_i \\ P_{k_i} \end{pmatrix} \qquad (13)$$

In Eq. (11), the elements of space group corresponding to a unit element of the permutation group $S_{k_n}$ form a group. This group is expressed as follows.

$$\cdots \cap G_{k_i} \cap G_{k_j} \cap G_{k_k} \cap \cdots = G_e \qquad (14)$$

$G_e$ is obviously invariant subgroup, and so, the channel group is decomposed as follows.

$$G_{[k]} = g_0 G_e + g_1 G_e + \cdots + g_n G_e \qquad (15)$$

Thus, we can define the factor group which the channel group $G_{[k]}$ is divided by $G_e$. That is,

$$G_f = G_{[k]} / G_e \qquad (16)$$

Factor group $G_f$ is generally isomorphic to the subgroup of $n$-dimensional permutation group $S_{k_n}$.

Since the star channel group is also a space group, we can also define the point group and the translation group of the star channel.

Star channel group $G_{[k]}$ has translation group $T_{[k]}$ which is an invariant subgroup. We call this group $T_{[k]}$ the translation group of star channel.

$G_{[k]}$ is decomposed to left cosets of $T_{[k]}$ as follows.

$$G_{[k]} = g_1 T_{[k]} + g_2 T_{[k]} + \cdots + g_s T_{[k]} \qquad (17)$$

Because $T_{[k]}$ is an invariant subgroup of $G_{[k]}$, we can define the factor group $G_{[k]} / T_{[k]}$. We call it the point group of the star channel ($G^0_{[k]}$).

Star channel group $G_{[k]}$ can be used to consider directly the actual symmetry of phase transition during various spontaneous symmetry breakings.

Suppose that the phase transition $G_0 \leftrightarrow G_1 \leftrightarrow G_2 \leftrightarrow \cdots \leftrightarrow G_n$ occurs in crystal. If $G_n$ is related to some arms of the given star, we can consider that lower symmetry phases $G_1$, $G_2$, …, $G_n$ are obtained from the star channel group $G_{[k]n}$ during the phase transition, that is, $G_{[k]n} \leftrightarrow G_1 \leftrightarrow G_2 \leftrightarrow \cdots \leftrightarrow G_n$.

Let us consider the symmetry breakings in $CsPbCl_3$ ferroelastic crystal and $Pb_{1-x}Ca_xTiO_3$(PCT) crystal discussed respectively in Refs. [15, 17] and Refs. [20-22].

In Refs. [15, 17], from the viewpoint of the theory of full irreducible representation of space group, it regards that the continuous phase transitions $O_h^1 \xrightarrow{319K} D_{4h}^5 \xrightarrow{317K} D_{2h}^{17} \xrightarrow{310K} D_{2h}^{16}(C_{2h}^2)$ are induced by the six-dimensional reducible representation of Lifshitz stars $k_{11}$ and $k_{13}$ of space group $O_h^1$. Also, in Refs. [20-22], from the same viewpoint, it regards that the phase transition in $Pb_{1-x}Ca_xTiO_3$(PCT) crystal are induced by the nine-dimensional reducible representation of Lifshitz reducible star $k_{11}+k_{12}+k_{13}$ of space group $O_h^1$ in the region 0<x<0.62.

As shown the Table 1, $k_{13}$ and $k_{12}$ have only an arm and $k_{11}$ has three arms. It was already experimentally known that, with respect to the star $k_{11}$, the above phase transitions are related to only an arm but no all arms of it.[22] Therefore, there is the possibility to use the wave vector star channel groups related with some arms taking part in the phase transitions.

The possible star channels of the reducible stars [$k_{11}+k_{13}$] and [$k_{11}+k_{12}+k_{13}$] are shown the Table 2, where the number in brackets with the star channel, e. g., [$k_{12}$](1), denotes the volume change of unite cell during the phase transition. We can see that in the reducible star channels, not all arbitrary combinations of the irreducible star channels, but only some selected combinations permitted by the symmetry of parent phase are contained.

Analyzing the lattice types and the point groups according to each star channel, we can know that the phase transition in $CsPbCl_3$ ferroelastic crystal is described by the group of star channel $[k_{11}^{(i)} + k_{13}](i=1, 2, 3)$ with the point group $D_{4h}$. That is, it can be supposed that this phase transition is occurred from the parent phase with the point group $D_{4h}$ but no the space group $O_h^1$. Similarly, we can also know that the phase transition in $Pb_{1-x}Ca_xTiO_3$(PCT) crystal can be described by the group of star channel $[k_{11}^{(i)} + k_{12} + k_{13}]$ $(i=1, 2, 3)$ with the point group $D_{4h}$. Then, the orders and the dimensions of the corresponding groups become very lower, and so, we can use the new potential models with the bases of which the numbers of invariants are also pretty decreased. Therefore, we can effectively study both the symmetry and the phenomenological theory for these crystals.

## 4. Conclusion

In this paper, we suggested a kind of the new method for determining the wave vector star channel in close relation with symmetry of order parameter. Also, for the first time we defined the wave vector star channel group, which can be efficiently used in studying the translational symmetry breaking related to all the Lifshitz wave vector stars of 230 space groups.

By accepting the new order parameter space spanned by the representation of translation group and finding out the subspaces invariant to the image of the representation, we established a method to find out the wave vector star channels. We, using the applied examples, showed that the wave vector star channel group is very effective in studying the translational symmetry breaking.

**Table 1**
Star Channels of Lifshitz wave vector stars of 230 space groups

**Table 2**
Star channels, types of lattice and point groups of the reducible stars [$k_{11}+k_{13}$] and [$k_{11}+k_{12}+k_{13}$]

Triclinic ($\Gamma_t$)

| star | Ref. [3] channel | n | This work [$k$] | star | Ref. [3] channel | n | This work [$k$] |
|---|---|---|---|---|---|---|---|
| $k_1 = \frac{1}{2}(b_1+b_2+b_3)$ | $k_1$ | 1 |  | $k_5 = \frac{1}{2}b_1$ | $k_5$ |  |  |
| $k_2 = \frac{1}{2}(b_2+b_3)$ | $k_2$ |  | [$k$] | $k_6 = \frac{1}{2}b_2$ | $k_6$ | 2 | [$k$] |
| $k_3 = \frac{1}{2}(b_1+b_3)$ | $k_3$ | 2 |  | $k_7 = \frac{1}{2}b_3$ | $k_7$ |  |  |
| $k_4 = \frac{1}{2}(b_1+b_2)$ | $k_4$ |  |  | $k_8 = 0$ | $k_8$ |  |  |

Monoclinic P ($\Gamma_m$)

| star | Ref. [3] channel | n | This work [$k$] | star | Ref. [3] channel | n | This work [$k$] |
|---|---|---|---|---|---|---|---|
| $k_7 = 0$ | $k_7$ | 1 |  | $k_{11} = \frac{1}{2}b_3$ | $k_{11}$ |  |  |
| $k_8 = \frac{1}{2}(b_1+b_3)$ | $k_8$ |  | [$k$] | $k_{12} = \frac{1}{2}b_1$ | $k_{12}$ | 2 | [$k$] |
| $k_9 = \frac{1}{2}(b_2+b_3)$ | $k_9$ | 2 |  | $k_{13} = \frac{1}{2}b_2$ | $k_{13}$ |  |  |
| $k_{10} = \frac{1}{2}(b_1+b_2+b_3)$ | $k_{10}$ |  |  | $k_{14} = \frac{1}{2}(b_1+b_2)$ | $k_{14}$ |  |  |

Monoclinic A ($\Gamma_m^b$)

| star | Ref. [3] channel | n | This work [$k$] | star | Ref. [3] channel | n | This work [$k$] |
|---|---|---|---|---|---|---|---|
| $k_6 = 0$ | $k_6$ | 1 |  | $k_4^{(1)} = \frac{1}{2}b_2$ | (i) | 2 | [$k_4^{(1)}$], [$k_4^{(2)}$] |
| $k_7 = \frac{1}{2}b_1$ | $k_7$ |  | [$k$] | $k_4^{(2)} = -\frac{1}{2}b_3$ | (12) | 4 | [$k_4^{(12)}$] |
| $k_8 = \frac{1}{2}(b_2+b_3)$ | $k_8$ | 2 |  | $k_5^{(1)} = \frac{1}{2}(b_1+b_2)$ | (i) | 2 | [$k_5^{(1)}$], [$k_5^{(2)}$] |
| $k_9 = \frac{1}{2}(b_1+b_2+b_3)$ | $k_9$ |  |  | $k_5^{(2)} = -\frac{1}{2}(b_1+b_3)$ | (12) | 4 | [$k_5^{(12)}$] |

### Orthorhombic P ($\Gamma_o$)

| star | Ref. [3] | | This work | star | Ref. [3] | | This work |
|---|---|---|---|---|---|---|---|
| | channel | n | [$k$] | | channel | n | [$k$] |
| $k_{19}=0$ | $k_{19}$ | 1 | | $k_{23}=\frac{1}{2}(b_2+b_3)$ | $k_{23}$ | | |
| $k_{20}=\frac{1}{2}b_1$ | $k_{20}$ | | [$k$] | $k_{24}=\frac{1}{2}(b_1+b_3)$ | $k_{24}$ | | [$k$] |
| $k_{21}=\frac{1}{2}b_2$ | $k_{21}$ | 2 | | $k_{25}=\frac{1}{2}(b_1+b_2)$ | $k_{25}$ | 2 | |
| $k_{22}=\frac{1}{2}b_3$ | $k_{22}$ | | | $k_{26}=\frac{1}{2}(b_1+b_2+b_3)$ | $k_{26}$ | | |

### Orthorhombic C ($\Gamma_o^b$)

| star | Ref. [3] | | This work | star | Ref. [3] | | This work |
|---|---|---|---|---|---|---|---|
| | channel | n | [$k$] | | channel | n | [$k$] |
| $k_{14}=0$ | $k_{14}$ | 1 | | $k_{12}^{(1)}=\frac{1}{2}b_1$ | ($i$) | 2 | [$k_{12}^{(1)}$], [$k_{12}^{(2)}$] |
| $k_{15}=\frac{1}{2}(b_1+b_2)$ | $k_{15}$ | | [$k$] | $k_{12}^{(2)}=-\frac{1}{2}b_2$ | (12) | 4 | [$k_{12}^{(12)}$] |
| $k_{16}=\frac{1}{2}b_3$ | $k_{16}$ | 2 | | $k_{13}^{(1)}=\frac{1}{2}(b_1+b_3)$ | ($i$) | 2 | [$k_{13}^{(1)}$], [$k_{13}^{(2)}$] |
| $k_{17}=\frac{1}{2}(b_1+b_2+b_3)$ | $k_{17}$ | | | $k_{13}^{(2)}=-\frac{1}{2}(b_2+b_3)$ | (12) | 4 | [$k_{13}^{(12)}$] |

### Orthorhombic F ($\Gamma_o^f$)

| star | Ref. [3] | | This work | star | Ref. [3] | | This work |
|---|---|---|---|---|---|---|---|
| | channel | n | [$k$] | | channel | n | [$k$] |
| $k_{14}=0$ | $k_{14}$ | 1 | | | ($i$) | 2 | [$k_{10}^{(1)}$], [$k_{10}^{(2)}$], [$k_{10}^{(3)}$], [$k_{10}^{(4)}$] |
| $k_{15}=\frac{1}{2}(b_2+b_3)$ | $k_{15}$ | | [$k$] | $k_{10}^{(1)}=\frac{1}{2}b_1$ $k_{10}^{(2)}=-\frac{1}{2}(b_1+b_2+b_3)$ $k_{10}^{(3)}=\frac{1}{2}b_3$ $k_{10}^{(4)}=\frac{1}{2}b_2$ | (12) | 4 | [$k_{10}^{(14)}$],[$k_{10}^{(23)}$], [$k_{10}^{(12)}$],[$k_{10}^{(34)}$], [$k_{10}^{(13)}$],[$k_{10}^{(24)}$] |
| $k_{16}=\frac{1}{2}(b_1+b_3)$ | $k_{16}$ | 2 | | | ($ijk$) | 8 | (*) |
| $k_{17}=\frac{1}{2}(b_1+b_2)$ | $k_{17}$ | | | | (1234) | | [$k_{10}^{(1234)}$] |

## Orthorhombic I ($\Gamma_o^v$)

| star | Ref. [3] | | This work | star | Ref. [3] | | This work |
|---|---|---|---|---|---|---|---|
| | channel | n | [$k$] | | channel | n | [$k$] |
| $k_{17}=0$ | $k_{17}$ | 1 | [$k$] | $k_{15}^{(1)} = \frac{1}{2} b_3$ | ($i$) | 2 | [$k_{15}^{(1)}$], [$k_{15}^{(2)}$] |
| $k_{18}= \frac{1}{2} (b_1+b_2+b_3)$ | $k_{18}$ | 2 | | $k_{15}^{(2)} = \frac{1}{2} ( b_2-b_1)$ | (12) | 4 | [$k_{15}^{(12)}$] |
| $k_{13}^{(1)} = \frac{1}{2} b_1$ | ($i$) | 2 | [$k_{13}^{(1)}$], [$k_{13}^{(2)}$] | $k_{16}^{(1)} = \frac{1}{4} (b_1+b_2+b_3)$ | ($i$) | 4 | (*) |
| $k_{13}^{(2)} = \frac{1}{2} ( b_2-b_3)$ | (12) | 4 | [$k_{13}^{(12)}$] | $k_{16}^{(2)} = - k_{16}^{(1)}$ | (12) | | [$k_{16}^{(12)}$] |
| $k_{14}^{(1)} = \frac{1}{2} b_2$ | ($i$) | 2 | [$k_{14}^{(1)}$], [$k_{14}^{(2)}$] | | | | |
| $k_{14}^{(2)} = \frac{1}{2} ( b_1-b_3)$ | (12) | 4 | [$k_{14}^{(12)}$] | | | | |

## Tetragonal P ($\Gamma_q$)

| star | Ref. [3] | | This work | star | Ref. [3] | | This work |
|---|---|---|---|---|---|---|---|
| | channel | n | [$k$] | | channel | n | [$k$] |
| $k_{17}=0$ | $k_{17}$ | 1 | [$k$] | $k_{15}^{(1)} = \frac{1}{2} b_2$ | ($i$) | 2 | [$k_{15}^{(1)}$], [$k_{15}^{(2)}$] |
| $k_{18}= \frac{1}{2} (b_1+b_2)$ | $k_{18}$ | | | $k_{15}^{(2)} = -\frac{1}{2} b_1$ | (12) | 2 | [$k_{15}^{(12)}$] |
| $k_{19}= \frac{1}{2} b_3$ | $k_{19}$ | 2 | | $k_{16}^{(1)} = \frac{1}{2} ( b_2+b_3)$ | ($i$) | 2 | [$k_{16}^{(1)}$], [$k_{16}^{(2)}$] |
| $k_{20}= \frac{1}{2} (b_1+b_2+b_3)$ | $k_{20}$ | | | $k_{16}^{(2)} = -\frac{1}{2} ( b_1+b_2)$ | (12) | 2 | [$k_{16}^{(12)}$] |

## Tetragonal I ($\Gamma_q^v$)

| star | Ref. [3] | | This work | star | Ref. [3] | | This work |
|---|---|---|---|---|---|---|---|
| | channel | n | [$k$] | | channel | n | [$k$] |
| $k_{14}=0$ | $k_{14}$ | 1 | [$k$] | $k_{11}^{(1)} = \frac{1}{2} b_2$ | ($i$) | 2 | [$k_{11}^{(1)}$], [$k_{11}^{(2)}$], [$k_{11}^{(3)}$], [$k_{11}^{(4)}$] |
| $k_{15}= \frac{1}{2} (b_1+b_2+b_3)$ | $k_{15}$ | 2 | | $k_{11}^{(2)} = \frac{1}{2} b_1$ | (13), (24) | 2 | [$k_{11}^{(13)}$], [$k_{11}^{(24)}$] |
| $k_{13}^{(1)} = \frac{1}{2} b_3$ | ($i$) | 2 | [$k_{13}^{(1)}$], [$k_{13}^{(2)}$] | $k_{11}^{(3)} = \frac{1}{2} ( b_3-b_1)$ | (12), (14), (23), (34) | 4 | [$k_{11}^{(12)}$], [$k_{11}^{(14)}$], [$k_{11}^{(23)}$], [$k_{11}^{(34)}$] |
| $k_{13}^{(2)} = \frac{1}{2} ( b_1-b_2)$ | (12) | 4 | [$k_{13}^{(12)}$] | $k_{11}^{(4)} = \frac{1}{2} ( b_3-b_2)$ | | | |
| $k_{12}^{(1)} = \frac{1}{4} (b_1+b_2+b_3)$ | ($i$) | 4 | (*) | | ($ijk$) | 8 | (*) |
| $k_{12}^{(2)} = - k_{12}^{(1)}$ | (12) | | [$k_{12}^{(12)}$] | | (1234) | | [$k_{11}^{(1234)}$] |

## Rhombohedral P ($\Gamma_{rh}$)

| star | Ref. [3] channel | n | This work [k] | star | Ref. [3] channel | n | This work [k] |
|---|---|---|---|---|---|---|---|
| $k_7=0$ | $k_7$ | 1 | [k] | | (i) | 2 | $[k_4^{(1)}], [k_4^{(2)}], [k_4^{(3)}]$ |
| $k_8=\frac{1}{2}(b_1+b_2+b_3)$ | $k_8$ | 2 | | $k_4^{(1)}=\frac{1}{2}b_3$ | | | |
| $k_5^{(1)}=\frac{1}{2}(b_1+b_2)$ | (i) | 2 | $[k_5^{(1)}], [k_5^{(2)}], [k_5^{(3)}]$ | $k_4^{(2)}=\frac{1}{2}b_2$ | (ij) | 4 | $[k_4^{(12)}], [k_4^{(13)}], [k_4^{(23)}]$ |
| $k_5^{(2)}=\frac{1}{2}(b_1+b_3)$ | (ij) | 4 | (*) | $k_4^{(3)}=\frac{1}{2}b_1$ | | | |
| $k_5^{(3)}=\frac{1}{2}(b_2+b_3)$ | (123) | | $[k_5^{(123)}]$ | | (123) | 8 | $[k_4^{(123)}]$ |

## Hexagonal P ($\Gamma_h$)

| star | Ref. [3] channel | n | This work [k] | star | Ref. [3] channel | n | This work [k] |
|---|---|---|---|---|---|---|---|
| $k_{16}=0$ | $k_{16}$ | 1 | [k] | | (i) | 2 | $[k_{14}^{(1)}], [k_{14}^{(1)}], [k_{14}^{(1)}]$ |
| $k_{17}=\frac{1}{2}(b_1+b_2+b_3)$ | $k_{17}$ | 2 | | $k_{14}^{(1)}=\frac{1}{2}(b_1+b_3)$ | | | |
| $k_{12}^{(1)}=\frac{1}{2}b_1$ | (i) | 2 | $[k_{12}^{(1)}], [k_{12}^{(2)}], [k_{12}^{(3)}]$ | $k_{14}^{(2)}=\frac{1}{2}(b_2+b_3)$ | (ij) | 4 | $[k_{14}^{(12)}], [k_{14}^{(13)}], [k_{14}^{(23)}]$ |
| $k_{12}^{(2)}=\frac{1}{2}b_2$ | (ij) | 4 | (*) | $k_{14}^{(3)}=-\frac{1}{2}(b_1+b_2-b_3)$ | | | |
| $k_{12}^{(3)}=-\frac{1}{2}(b_1+b_2)$ | (123) | | $[k_{12}^{(123)}]$ | | (123) | 8 | $[k_{14}^{(123)}]$ |
| $k_{13}^{(1)}=\frac{1}{3}(b_1+b_2)$ | (i) | 3 | (*) | $k_{15}^{(1)}=\frac{1}{3}(b_1+b_2)+\frac{1}{2}b_3$ | (i) | 6 | (*) |
| $k_{13}^{(2)}=-k_{13}^{(1)}$ | (12) | | $[k_{13}^{(12)}]$ | $k_{15}^{(2)}=-k_{15}^{(1)}$ | (12) | | $[k_{15}^{(12)}]$ |

## Cubic P ($\Gamma_c$)

| star | Ref. [3] channel | n | This work [k] | star | Ref. [3] channel | n | This work [k] |
|---|---|---|---|---|---|---|---|
| $k_{12}=0$ | $k_{12}$ | 1 | [k] | | (i) | 2 | $[k_{10}^{(1)}], [k_{10}^{(2)}], [k_{10}^{(3)}]$ |
| $k_{13}=\frac{1}{2}(b_1+b_2+b_3)$ | $k_{13}$ | 2 | | $k_{10}^{(1)}=\frac{1}{2}b_3$ | | | |
| $k_{11}^{(1)}=\frac{1}{2}(b_1+b_2)$ | (i) | 2 | $[k_{11}^{(1)}], [k_{11}^{(2)}], [k_{11}^{(3)}]$ | $k_{10}^{(2)}=\frac{1}{2}b_2$ | (ij) | 4 | $[k_{10}^{(12)}], [k_{10}^{(13)}], [k_{10}^{(23)}]$ |
| $k_{11}^{(2)}=\frac{1}{2}(b_1+b_3)$ | (ij) | 4 | (*) | $k_{10}^{(3)}=\frac{1}{2}b_1$ | | | |
| $k_{11}^{(3)}=\frac{1}{2}(b_2+b_3)$ | (123) | | $[k_{11}^{(123)}]$ | | (123) | 8 | $[k_{10}^{(123)}]$ |

## Cubic I ($\Gamma_c^v$)

| star | Ref. [3] channel | n | This work [k] | star | Ref. [3] channel | n | This work [k] |
|---|---|---|---|---|---|---|---|
| $k_{11}=0$ | $k_{11}$ | 1 | [k] | | (i) | 2 | $[k_9^{(1)}],[k_9^{(2)}],$ $[k_9^{(3)}],[k_9^{(4)}],$ $[k_9^{(5)}],[k_9^{(6)}]$ |
| $k_{12}=\frac{1}{2}(b_1+b_2-b_3)$ | $k_{12}$ | 2 | | $k_9^{(1)}=\frac{1}{2}b_3$ $k_9^{(2)}=\frac{1}{2}b_2$ | (14), (25), (36) | 4 | $[k_9^{(14)}],$ $[k_9^{(25)}],[k_9^{(36)}]$ |
| $k_{10}^{(1)}=\frac{1}{4}(b_1+b_2+b_3),$ $k_{10}^{(2)}=-k_{10}^{(1)}$ | (i) | 4 | (*) | $k_9^{(3)}=\frac{1}{2}b_1$ $k_9^{(4)}=\frac{1}{2}(b_2-b_1)$ | (126), (135), (234), (456) | 4 | $[k_9^{(126)}],[k_9^{(135)}],$ $[k_9^{(234)}],[k_9^{(456)}]$ |
| | (12) | | $k_{10}^{(12)}$ | $k_9^{(5)}=\frac{1}{2}(b_3-b_1)$ $k_9^{(6)}=\frac{1}{2}(b_3-b_2)$ | (ij) (except (14), (25), (36)) | 4 | (*) |
| $k_9^{(1\sim6)}$ | (ijkl) | 8 | (*) | | (ijk) (except (126), (135), (234), (456)) | 8 | (*) |
| | (ijklm) | | (*) | | | | |
| | (1~6) | | $[k_9^{(1\sim6)}]$ | | | | |

## Cubic F ($\Gamma_c^f$)

| star | Ref. [3] channel | n | This work [k] | star | Ref. [3] channel | n | This work [k] |
|---|---|---|---|---|---|---|---|
| $k_{11}=0$ | $k_{11}$ | 1 | [k] | | (13), (14), (15), (16), (23), (24), (25), (26), (35), (36), (45), (46) | 16 | (*) |
| $k_{10}^{(1)}=\frac{1}{2}(b_1+b_2)$ $k_{10}^{(2)}=\frac{1}{2}(b_1+b_3)$ $k_{10}^{(3)}=\frac{1}{2}(b_2+b_3)$ | (i) | 2 | $[k_{10}^{(1)}],[k_{10}^{(2)}],$ $[k_{10}^{(3)}]$ | $k_8^{(1)}=\frac{1}{4}(b_1+b_2)+$ $\frac{1}{2}(b_2+b_3)$ $k_8^{(2)}=-k_8^{(1)}$ | (123), (124), (125), (126), (134), (156), (234), (256), (345), (346), (356), (456) | 16 | (*) |
| | (ij) | 4 | (*) | | | | |
| | (123) | | $[k_{10}^{(123)}]$ | $k_8^{(3)}=\frac{1}{4}(b_1+b_3)+$ $\frac{1}{2}(b_1+b_2)$ $k_8^{(4)}=-k_8^{(3)}$ | (135), (136), (145), (146), (235), (236), (245), (246) | 32 | (*) |
| $k_9^{(1)}=\frac{1}{2}(b_1+b_2+b_3)$ $k_9^{(2)}=\frac{1}{2}b_1$ $k_9^{(3)}=\frac{1}{2}b_2$ $k_9^{(4)}=\frac{1}{2}b_3$ | (i) | 2 | $[k_9^{(1)}],[k_9^{(2)}],$ $[k_9^{(3)}],[k_9^{(4)}]$ | | | | |
| | (ij) | 4 | $[k_9^{(12)}],[k_9^{(13)}],$ $[k_9^{(14)}],[k_9^{(23)}],$ $[k_9^{(24)}],[k_9^{(34)}]$ | $k_8^{(5)}=\frac{1}{4}(b_2+b_3)+$ $\frac{1}{2}(b_1+b_3)$ $k_8^{(6)}=-k_8^{(5)}$ | (1234), (1256), (3456) | 16 | $[k_8^{(1234)}],$ $[k_8^{(1256)}],$ $[k_8^{(3456)}]$ |
| | (ijk) | 8 | (*) | | | | |
| | (1~4) | | $[k_9^{(1234)}]$ | | | | |
| $k_8^{(1\sim6)}$ | (i) | 4 | (*) | | (ijkl) (except (1234), (1256), (3456)) | 32 | (*) |
| | (12), (34), (56) | 4 | $[k_8^{(12)}],[k_8^{(34)}],$ $[k_8^{(56)}]$ | | | | |
| | (ijklm) | 32 | (*) | | (123456) | 32 | $[k_8^{(123456)}]$ |

Table 2

Star channels, types of lattice and point groups of the reducible stars [$k_{11}+k_{13}$] and [$k_{11}+k_{12}+k_{13}$]

| Star channels of M+R | Star channels of M+R+Γ | Lattice types | Point groups |
|---|---|---|---|
| [$k_{11}^{(123)}$] (4) | [$k_{12}$](1) <br> [$k_{11}^{(123)}+k_{12}$](4) | $\Gamma_t$, $\Gamma_m$, $\Gamma_m^b$, $\Gamma_o$, $\Gamma_o^b$, $\Gamma_q$, $\Gamma_{rh}$, $\Gamma_c$ | $O_h$ |
| [$k_{11}^{(i)}$](2) <br> ($i=1, 2, 3$) | [$k_{11}^{(i)}+k_{12}$](2) <br> ($i=1, 2, 3$) | $\Gamma_t$, $\Gamma_m$, $\Gamma_m^b$, $\Gamma_o$, $\Gamma_o^b$, $\Gamma_q$ | $D_{4h}$ |
| [$k_{13}$](2) | [$k_{12}+k_{13}$](2) | $\Gamma_t$, $\Gamma_m^b$, $\Gamma_o^f$, $\Gamma_o^v$, $\Gamma_q^v$, $\Gamma_{rh}$, $\Gamma_c$ | $O_h$ |
| [$k_{11}^{(i)}+k_{13}$](4) <br> ($i=1, 2, 3$) | [$k_{11}^{(i)}+k_{12}+k_{13}$](4) <br> ($i=1, 2, 3$) | $\Gamma_t$, $\Gamma_m$, $\Gamma_m^b$, $\Gamma_o$, $\Gamma_o^b$, $\Gamma_q$ | $D_{4h}$ |